\documentclass[aps,prfluids,10pt,a4paper,onecolumn,superscriptaddress,balancelastpage,showpacs,notitlepage]{revtex4-2}

\usepackage[english]{babel}
\usepackage[T1]{fontenc}
\usepackage{endnotes}
\usepackage{amssymb,amsmath,amsfonts}
\usepackage{textcomp}
\usepackage{graphics,color}
\usepackage[colorlinks,citecolor=blue,urlcolor=blue,bookmarks=false,hypertexnames=true]{hyperref}

\usepackage{graphicx}

\usepackage{float}
\usepackage[export]{adjustbox}
\usepackage[abs]{overpic}
\usepackage{xcolor,varwidth}
\usepackage{layouts}
\usepackage{centernot}
\usepackage{bm}
\usepackage{amsmath}
\usepackage{times}
\usepackage{amssymb}
\usepackage{mathrsfs}
\usepackage{url}
\usepackage{version}
\usepackage{siunitx}
\usepackage{balance}
\usepackage{lastpage}
\usepackage{parskip}
\usepackage{siunitx}
\usepackage[paper=a4paper,margin=1in]{geometry}
\usepackage[caption=false]{subfig}

\usepackage{caption}
\definecolor{linkcolor}{rgb}{0,0,0.6}

\DeclareMathOperator\erf{erf}
\DeclareMathOperator\erfc{erfc}

\begin{document}

\title{Ewald summation of doubly periodic $2$- and $3$-dimensional Stokes multipoles}

\author{Teunike A. van de Pol}
\email{t.a.v.d.pol@tue.nl}
\affiliation{Division of Physical Chemistry, Lund University, Sweden}
\affiliation{Department of Applied Physics, Eindhoven University of Technology, P.O. Box 513, 5600 MB Eindhoven, The Netherlands}

\author{Zhan Ma}
\affiliation{School of Physics and Astronomy, The University of Edinburgh, James Clerk Maxwell Building, Peter Guthrie Tait Road, Edinburgh, EH9 3FD, United Kingdom}

\author{Alexander Morozov}
\affiliation{School of Physics and Astronomy, The University of Edinburgh, James Clerk Maxwell Building, Peter Guthrie Tait Road, Edinburgh, EH9 3FD, United Kingdom}

\author{Joakim Stenhammar}
\email{joakim.stenhammar@chem.lu.se}
\affiliation{Division of Physical Chemistry, Lund University, Sweden}

\date{\today}

\begin{abstract}
\noindent 
In spite of the qualitative differences between Stokes flows in $2$ and $3$ dimensions, most particle-resolved computational studies of microswimmers and other colloidal systems dominated by low-Re hydrodynamic interactions are carried out in triply periodic geometries, while the corresponding continuum descriptions are frequently solved in $2$-dimensional doubly periodic domains. Furthermore, most experimental examples of biological and synthetic microswimmers are realised in confined, quasi-2-dimensional geometries with a finite thickness in one of the three dimensions. To enable studying the non-trivial differences between these system geometries, in this paper we systematically derive Ewald representations for periodic Stokes force multipoles in geometries with two periodic directions, treating both three-dimensional Stokes flow with one unbounded coordinate and intrinsically two-dimensional singularities. Starting from the periodic Stokeslet, we obtain explicit expressions for force dipoles and quadrupoles and analyse the non-trivial zero-wavevector contribution arising under partial periodicity. The analytical expressions are independently validated by comparing with results from an explicit lattice Boltzmann solver in each geometry. 
\end{abstract}

\maketitle

\section{Introduction}

\noindent Suspensions of biological microswimmers such as bacteria or algae constitute archetypical examples of active matter dominated by long-ranged hydrodynamic interactions (HIs). These fluid-mediated interactions are central to both the interaction between individual swimmers and between swimmers and passive objects such as nutrients, manifesting themselves in long-ranged swimmer correlations~\citep{Stenhammar2017Correlations}, enhanced diffusion of passive tracers~\citep{Ishikawa2025Review,Mino2013} and bacterial collective motion~\citep{Dunkel2013ActTurb,BardfalvySoftMatter}. The fact that microswimmers live in an effectively overdamped environment allows for an efficient description of complex flow configurations in terms of force, torque and source multipoles~\citep{Spagnolie2012Quad}. For swimming bacteria such as \textit{E. coli}, a far-field description using a force dipole flow has proven very successful in describing both the enhanced transport of passive inclusions in bacterial suspensions~\cite{Morozov2014TracerDiff,Mino2013} and the transition to bacterial turbulence in rear-actuated (pusher) bacteria~\citep{Saintillan2008,Stenhammar2017Correlations}. The long-ranged nature of these flow fields, with the 3-dimensional ($3d$) force dipole decaying as $1/r^2$, poses challenges for particle-based computational descriptions, which typically employ $3d$ systems with triply periodic boundary conditions (PBCs). In contrast, many continuum descriptions of microswimmer suspensions, based either on explicitly derived kinetic theories~\citep{Ohm_Shelley_2022} or on phenomenological descriptions such as active nematics~\citep{Doostmohammadi2018}, are formulated on $2$-dimensional ($2d$), doubly periodic domains, where the hydrodynamic singularities decay slower than in $3d$. Adding to this complexity, in experimental settings microswimmer suspensions are usually realised in confined geometries such as thin films or near solid surfaces \cite{Wu2000Enhanced, Mathijssen2016Films, Aguayo2024Quasi2D}. Computationally, such geometries can be approximated by a quasi-$2d$ system with swimmers confined to a plane with two periodic directions~\citep{Bardfalvy2024Sheet}, while the $3d$ flow fields extend into the confined (non-periodic) third dimension, usually bounded by no-slip walls.

\begin{figure}[htbp]
    \centering
    \includegraphics[width=0.85\textwidth]{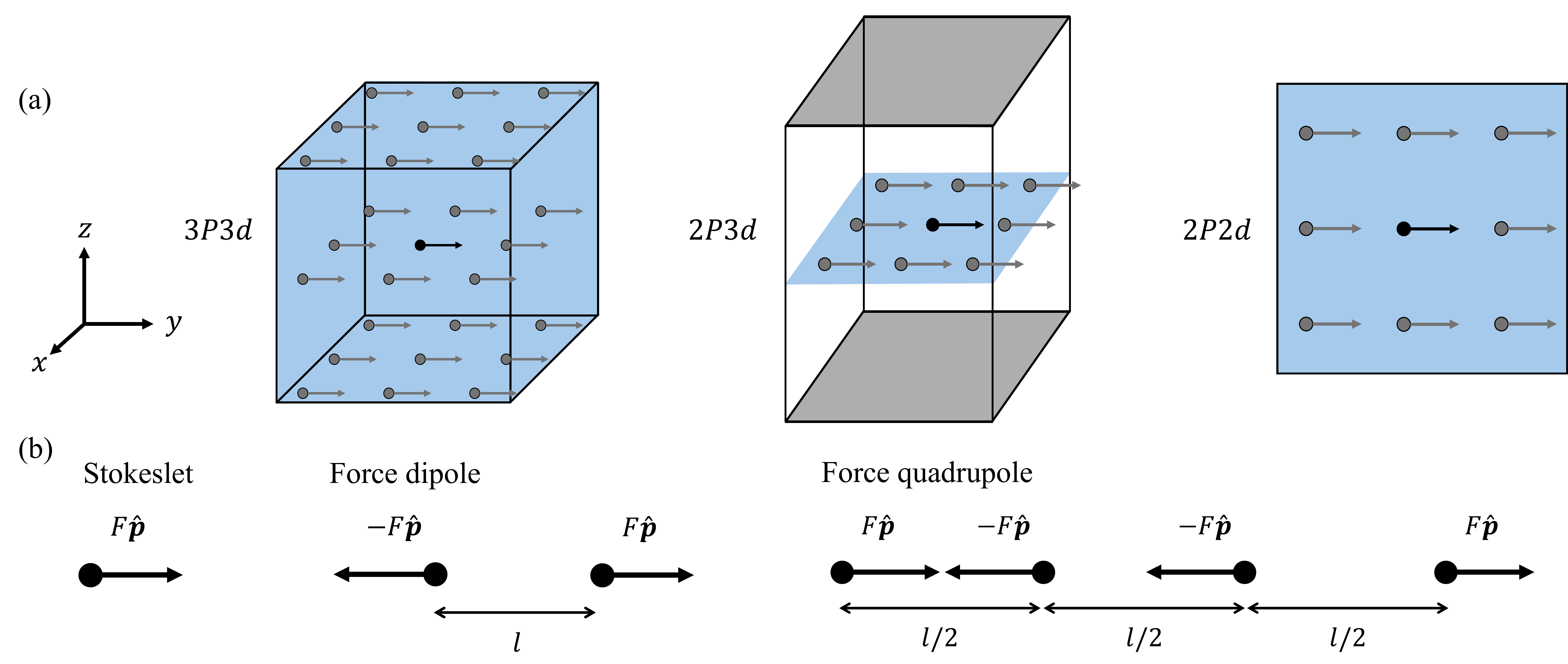}
    \caption{Schematic illustration of the geometries and force multipoles considered in this work. (a) Schematic of the three geometries considered in this work. In the doubly periodic $2P3d$ geometry, the system is periodic in two directions ($xy$), with the flow field and multipole orientation extending also into the third ($z$) direction. (b) Stokeslet and axisymmetric extended force dipole and quadrupole with orientation $F\hat{\bm{p}}$ and length $l$, as implemented in LB simulations. In the Ewald expressions we instead consider the point multipole limit $l \rightarrow 0$, $F \rightarrow \infty$ such that $Fl = \mathrm{const.}$ (dipole) or $Fl^2 = \mathrm{const.}$ (quadrupole)
    }
    \label{fig: array-geometry}
\end{figure}

Regardless of the geometry, using PBCs in systems with long-ranged interactions requires efficient computational handling of the resulting conditionally convergent lattice sums. For hydrodynamic solvers based on the summation of pairwise forces, the various incarnations of the Ewald summation technique provide an optimised framework for handling this. Originally developed for treating electrostatic interactions~\citep{Ewald1921Sum}, the Ewald method relies on the splitting of the infinite lattice sum into short- and long-ranged parts that are summed, respectively, in real and reciprocal space, leading to an absolutely convergent sum with dramatically faster convergence~\citep{Toukmaji1996RevEwald, Pozrikidis1996Periodic, afKlinteberg2014Ewald, deGraaf2017Ewald}. Due to the linear nature of the Stokes equation, Ewald summation is also ideally applied to the pairwise summation of Stokes flows. In this context, it has been used to describe flow fields and hydrodynamic interactions due to triply periodic arrays of 3-dimensional force singularities (``$3P3d$'') \cite{Hasimoto1959Ewald3D, Beenakker1986Ewald, Pozrikidis1996Periodic} and doubly periodic arrays of both 3-dimensional (``$2P3d$'')~\citep{Pozrikidis1996Periodic, Bleibel2012Quasi2D, Lindbo2011QuasiStokes} and 2-dimensional (``$2P2d$'')~\citep{vandeVorst1996Ewald2D} force singularities, as summarised in Fig.~\ref{fig: array-geometry}a. In the $3P3d$ setting, the Ewald-summed point-force solutions have also been used to derive higher-order force~\cite{deGraaf2017Ewald} and torque~\citep{afKlinteberg2014Ewald} multipoles, relevant for microswimmer flows, as well as spherical multipoles for the description of surface-slip driven active colloids  \citep{Deb2025EwaldColloid}.  

In this work, we build on previous formulations of the periodic Stokeslet singularity in $2P2d$~\citep{vandeVorst1996Ewald2D} and $2P3d$~\citep{Pozrikidis1996Periodic,Lindbo2011QuasiStokes,Bleibel2012Quasi2D} geometries, whose derivations we reproduce in slightly different formulations and verify. From these results, we derive analytical expressions for the higher-order flow fields generated by periodic arrays of force dipoles and quadrupoles (Fig.~\ref{fig: array-geometry}b) in $2P3d$ and $2P2d$ geometries, including subtleties arising from the long-wavelength corrections to the far-field dipole flow in the $2P3d$ setting. We independently verify our results by quantitative comparison of all the generated flow fields with numerical results using a point-force implementation of the lattice Boltzmann (LB) method in the three relevant geometries, including the necessary force cancellation needed for the Stokeslet singularities. Since the LB method solves the Stokes equation directly on a periodic domain, it does not require the explicit evaluation of lattice sums, and thus provides an independent and accurate verification of the derived expressions.  

The paper is organised as follows: We first introduce the Ewald summation framework and the relevant force multipoles in Section~\ref{sec: theory}. In Section~\ref{sec: 3d} we review the previously established $3P3d$ Ewald expressions which serve as a reference for our derivation, and additionally derive novel expressions for the $3P3d$ quadrupole flow. In Section~\ref{sec: q2d} we derive Ewald expressions for the Stokeslet, dipole and quadrupole in the $2P3d$ geometry, and in Section~\ref{sec: 2d} we derive the corresponding $2P2d$ expressions. In Section~\ref{sec: numerical} we finally provide numerical verification of our expressions, with a special focus on discussing the non-trivial $2P3d$ case. 

\section{Basic properties of periodic Stokes flows} \label{sec: theory}

\noindent Our starting point is the fundamental solution of the Stokes equations under point forcing, \textit{i.e.},
\begin{equation} \label{eq: Stokes-eq-delta}
    -\nabla p(\bm{r}) + \mu \nabla^2 \bm{u}(\bm{r}) = -\bm{F} \delta(\bm{r}),\quad \nabla \cdot \bm{u}(\bm{r}) = 0,
\end{equation}
where $\bm{u}(\bm{r})$ is the fluid velocity, $p(\bm{r})$ is the pressure, $\mu$ is the dynamic viscosity, and the force $\bm{F}$ is decomposed as $\bm{F} = F\hat{\bm{p}}$ with $F = |\bm{F}|$. Hats are used throughout this work to denote unit vectors, \textit{i.e.}, $\hat{\bm{a}} = \bm{a} / |\bm{a}|$, while the juxtaposition of two vectors denotes a dyadic product, \textit{i.e.}, $(\bm{a} \bm{b})_{ij} = a_i b_j$. The solution to Eq.~\eqref{eq: Stokes-eq-delta} is referred to as the Stokeslet and can be written in $d$ dimensions as
\begin{equation} \label{eq: u(r)-gen}
    \bm{u}^{(0)}_{d}(\bm{r}) = \frac{F}{8\pi\mu} \mathbf{S}_d(\bm{r}) \cdot \hat{\bm{p}},
\end{equation}
where the $d$-dimensional Green's function $\mathbf{S}_d(\bm{r})$, commonly referred to as the Oseen tensor, is given by
\begin{equation} \label{eq: S(r)-gen}
    \mathbf{S}_d(\bm{r}) = \big( \mathbb{I} \nabla^2 - \nabla \nabla \big) B_d(r).
\end{equation}
Here, $B_d(r)$ is the solution to the $d$-dimensional biharmonic equation, with
\begin{equation}
\begin{aligned}
    B_{2d}(r) &= r^2 - r^2 \ln (r) \\
    B_{3d}(r) &= r.
\end{aligned}
\end{equation}
We now consider a periodic array of identical point forces, separated by a distance $L$ in each spatial direction, where the central unit cell extends from $-L/2$ to $L/2$ and contains a single point force at its centre. Due to the linearity of the Stokes equation, we may write the flow field due to the array of forces as a sum of the individual flow fields:
\begin{equation} \label{eq: sum-periodic}
    \bm{u}^{(0)}_d(\bm{r}) = \frac{F}{8\pi\mu} \sum_{\bm{r}_n} \mathbf{S}_d(\bm{r} - \bm{r}_n)\cdot\hat{\bm{p}}.
\end{equation}
The summation runs over all periodic images, with $\bm{r}_n = L \big( n_x \hat{\bm{x}} + n_y \hat{\bm{y}} + n_z \hat{\bm{z}} \big)$ ($3P3d$) or $\bm{r}_n = L \big( n_x \hat{\bm{x}} + n_y \hat{\bm{y}} \big)$ ($2P3d$ and $2P2d$), where $n_x$, $n_y$ and $n_z$ are integers. To separate the Oseen tensor in short- and long-ranged contributions, we introduce a splitting function $g_d(\xi r)$, with $\xi$ a splitting parameter with dimensions of inverse length. By adding and subtracting the splitting function to $B_d(r)$ \cite{deGraaf2017Ewald}, \textit{i.e.},
\begin{equation}
    \mathbf{S}_d(\bm{r}) = (\mathbb{I}\nabla^2 - \nabla\nabla) \big[ B_d(r) + g_d(\xi r) - g_d(\xi r) \big] = \mathbf{\Theta}_d(\bm{r}) + \mathbf{\Psi}_d(\bm{r}),
\end{equation}
we can define the functions $\mathbf{\Theta}_d(\bm{r})$ and $\mathbf{\Psi}_d(\bm{r})$ as
\begin{equation} \label{eq:Theta_Phi}
\begin{aligned}
    \mathbf{\Theta}_d(\bm{r}) &\equiv  (\mathbb{I}\nabla^2 - \nabla\nabla) g_d(\xi r), \\
    \mathbf{\Psi}_d(\bm{r}) &\equiv  (\mathbb{I}\nabla^2 - \nabla\nabla) \big[ B_d(r) - g_d(\xi r) \big].
\end{aligned}
\end{equation}
The requirements for a successful splitting are (\textit{i}) that the function $g_d(\xi r)$ decays considerably faster than the corresponding ``bare'' Stokeslet, leading to a quickly decaying  $\mathbf{\Theta}_d$, and (\textit{ii}) that the long-range part $\mathbf{\Psi}_d$ is a smooth function which yields an expression that converges rapidly in Fourier space. Under these conditions, $\mathbf{\Theta}_d$ can be evaluated directly in real space, while $\mathbf{\Psi}_d$ is summed in Fourier space using Poisson's summation formula~\citep{Pozrikidis1996Periodic}:
\begin{equation}
    \sum_{\bm{r}_n} \bm{\Psi}_d(\bm{r} - \bm{r}_n) = \frac{1}{V} \sum_{\bm{k}} e^{i\bm{k}\cdot\bm{r}}\, \bm{\Psi}_d(\bm{k}),
\end{equation}
where $V = L^d$ is the volume of the unit cell and the Fourier space sum runs over all wavevectors $\bm{k} = \frac{2\pi}{L} (\lambda_x \hat{\bm{x}} + \lambda_y \hat{\bm{y}} + \lambda_z \hat{\bm{z}})$ ($3P3d$) or $\bm{k} = \frac{2\pi}{L} (\lambda_x \hat{\bm{x}} + \lambda_y \hat{\bm{y}})$ ($2P3d$ and $2P2d$), where $\lambda_x$, $\lambda_y$ and $\lambda_z$ are integers. The $d$-dimensional Fourier transform is furthermore defined as
\begin{equation}
\begin{aligned}
    \mathcal{F}_d\{f(\bm{r})\} &= \int \mathrm{d}^d \bm{r}\, e^{-i\bm{k}\cdot\bm{r}} f(\bm{r}).
\end{aligned}
\end{equation}
In this framework, we may therefore express the $d$-dimensional velocity field of a periodic array of Stokeslets $\bm{u}^{(0)}_{d}(\bm{r})$ as 
\begin{equation} \label{eq: u(r)-Ewald}
    \bm{u}^{(0)}_{d}(\bm{r}) = \frac{F}{8\pi\mu} \Big( \sum_{\bm{r}_n} \mathbf{\Theta}_d(\bm{r}-\bm{r}_n)\cdot\hat{\bm{p}} + \frac{1}{L^d}\sum_{\bm{k}} e^{i\bm{k} \cdot \bm{r}}\, \mathbf{\Psi}_d(\bm{k})\cdot\hat{\bm{p}} \Big).
\end{equation}
The Stokeslet velocity field can be used to derive higher-order singularities. In the limit $l \rightarrow 0$, the force dipole can be obtained as the directional derivative in the direction of the dipole orientation \cite{Lauga2020Book, deGraaf2017Ewald}:
\begin{equation} \label{eq: stresslet}
    \bm{u}_d^{(1)}(\bm{r})= -\frac{\kappa}{8\pi\mu} (\hat{\bm{p}}\cdot\nabla) \mathbf{S}_d(\bm{r})\cdot\hat{\bm{p}},
\end{equation}
with $\kappa = Fl$ the dipole strength. For the force quadrupole, we use the convention that the forces are separated by $\frac{1}{2}l$ and centered at the origin, such that their position can be written as $\pm \frac{1}{4}l\hat{\bm{p}}$ and $\pm \frac{3}{4}l\hat{\bm{p}}$ (Fig.~\ref{fig: array-geometry}b). For $l \rightarrow 0$, the quadrupole flow field can then be expressed as \cite{Mathijssen2016Films, Spagnolie2012Quad, Deb2025EwaldColloid}:
\begin{equation} \label{eq: quadrupole-gen}
    \bm{u}_d^{(2)}(\bm{r}) = \frac{\nu}{16\pi\mu} (\hat{\bm{p}}\cdot\nabla)^2\, \mathbf{S}_d(\bm{r})\cdot\hat{\bm{p}},
\end{equation}
with $\nu = F l^2$ the quadrupole strength.

\section{Triply periodic $3d$ multipoles} \label{sec: 3d}

\noindent The Ewald summation framework described in the previous Section can be used in different dimensionalities, starting from the $d$-dimensional solution of the biharmonic equation and by selecting an appropriate splitting function that yields the required fast decay of the real-space and Fourier-space terms. For reference, we begin by reviewing previously established Ewald expressions for the Stokeslet \cite{Pozrikidis1996Periodic} and force dipole \cite{deGraaf2017Ewald} in triply periodic systems, in addition to deriving the corresponding $3P3d$ expressions for the force quadrupole. 

Several choices for the splitting function $g_{d}(\xi r)$ have been proposed, varying in computational efficiency and complexity \cite{Ewald1921Sum, Hasimoto1959Ewald3D, Beenakker1986Ewald, afKlinteberg2025PSW}. Here, we use the splitting function due to Beenakker~\citep{Beenakker1986Ewald}, who proposed $g_{3d}(\xi r) = r \erfc(\xi r)$, where $\erfc(x) = 1-\erf(x)$ is the complementary error function, with 
\begin{equation}
    \erf(x) = \frac{2}{\sqrt{\pi}} \int_0^x \text{d}t\,e^{-t^2}.
\end{equation}
With this choice, the $3P3d$ periodic Stokeslet expression becomes
\begin{equation} \label{eq: Ewald-3D-Stokeslet}
\begin{aligned}
    \bm{u}^{(0)}_{3P3d}(\bm{r}) &= \sum_{\bm{r}_n} \bm{u}^{(0)}_{R,3P3d}(\bm{r}-\bm{r}_n) + \frac{1}{L^3}\sum_{\bm{k}\neq\bm{0}} \bm{u}^{(0)}_{F,3P3d}(\bm{k}) \cos(\bm{k}\cdot\bm{r}) \\
    \bm{u}^{(0)}_{R,3P3d}(\bm{r}) &= \frac{F}{8\pi\mu} \Big( C_{3P3d}(\xi r) \frac{\mathbb{I}}{r}+ D_{3P3d}(\xi r) \frac{\bm{r}\bm{r}}{r^3}\Big) \cdot\hat{\bm{p}} \\
    \bm{u}^{(0)}_{F,3P3d}(\bm{k}) &= \frac{F}{\mu k^4} e^{-\omega^2/4} \left( 1 + \frac{\omega^2}{4} + \frac{\omega^4}{8} \right)( \mathbb{I}k^2 - \bm{k}\bm{k} ) \cdot\hat{\bm{p}},
\end{aligned}
\end{equation}
with the dimensionless wavenumber $\omega \equiv k/\xi$ and auxiliary functions 
\begin{equation}\label{eq:C_D_3d}
\begin{aligned}
    C_{3P3d}(a) &= \erfc(a) + \frac{2a}{\sqrt{\pi}} e^{-a^2} \big( 2 a^2 - 3 \big) \\
    D_{3P3d}(a) &= \erfc(a) + \frac{2a}{\sqrt{\pi}} e^{-a^2} \big( 1 - 2a^2 \big).
\end{aligned}
\end{equation}
To avoid a divergence as $k \rightarrow 0$, the $\bm{k}=\bm{0}$ term is omitted from the Fourier-space sum in Eq.~\eqref{eq: Ewald-3D-Stokeslet}, which physically corresponds to describing the flow velocity relative to the mean flow induced by having a net force acting on the fluid volume~\citep{Pozrikidis1996Periodic, deGraaf2017Ewald, Brady1988EwaldSim}. As we verify below, this can be approximated computationally by adding a set of homogeneously distributed background forces to the simulation domain, which together cancel out the net force acting on the fluid arising from the primary singularity. 

The expression for a periodic array of force dipoles can be obtained by differentiation as described in the previous section, leading to \cite{deGraaf2017Ewald} 
\begin{equation} \label{eq: Ewald-3D-stresslet}
\begin{aligned}
    \bm{u}_{3P3d}^{(1)}(\bm{r}) &= \sum_{\bm{r}_n} \bm{u}^{(1)}_{R,3P3d}(\bm{r}-\bm{r}_n) + \frac{1}{L^3} \sum_{\bm{k}\neq\bm{0}} \bm{u}^{(1)}_{F,3P3d}(\bm{k}) \sin(\bm{k}\cdot \bm{r}) \\
    \bm{u}^{(1)}_{R,3P3d}(\bm{r}) &= \frac{\kappa}{8\pi\mu}\frac{1}{r^2} \Big( E_{3P3d}(\xi r) (\hat{\bm{p}}\cdot\hat{\bm{r}})^2 \hat{\bm{r}} + F_{3P3d}(\xi r) (\hat{\bm{p}}\cdot\hat{\bm{r}})\hat{\bm{p}} - D_{3P3d}(\xi r) \hat{\bm{r}} \Big) \\
    \bm{u}^{(1)}_{F,3P3d}(\bm{k}) &= \frac{\kappa}{\mu k^3} (\hat{\bm{p}}\cdot\hat{\bm{k}}) e^{-\omega^2/4} \left( 1 + \frac{\omega^2}{4} + \frac{\omega^4}{8} \right)(\mathbb{I}k^2 - \bm{k}\bm{k})\cdot\hat{\bm{p}},
\end{aligned}
\end{equation}
with auxiliary functions 
\begin{equation}
    E_{3P3d}(a) = 3 \erfc(a) + \frac{2 a}{\sqrt{\pi}} e^{-a^2} \big( 3 + 2a^2 - 4a^4 \big)
\end{equation}
and
\begin{equation}
    F_{3P3d}(a) = \frac{2 a}{\sqrt{\pi}} e^{-a^2} \big( 4a^4 - 8a^2 \big),
\end{equation}
and $D_{3P3d}(a)$ is defined as for the Stokeslet (Eq.~\eqref{eq:C_D_3d}). The triply periodic force quadrupole similarly becomes 
\begin{equation} \label{eq: Ewald-3D-quadrupole}
\begin{aligned}
    \bm{u}_{3P3d}^{(2)}(\bm{r}) &= \sum_{\bm{r}_n} \bm{u}^{(2)}_{R,3P3d}(\bm{r}-\bm{r}_n)+ \frac{1}{L^3} \sum_{\bm{k}\neq \bm{0}} \bm{u}^{(2)}_{F,3P3d}(\bm{k}) \cos(\bm{k}\cdot\bm{r}) \\
    \bm{u}^{(2)}_{R,3P3d}(\bm{r}) &= \frac{\nu}{16\pi\mu}\frac{1}{r^3} \Big( H_{3P3d}(\xi r) \hat{\bm{p}} + I_{3P3d}(\xi r) (\hat{\bm{p}}\cdot\hat{\bm{r}})\hat{\bm{r}} 
    \\& \quad\quad\quad\quad\quad\quad\quad + J_{3P3d}(\xi r) (\hat{\bm{p}}\cdot\hat{\bm{r}})^2\hat{\bm{p}} + K_{3P3d}(\xi r) (\hat{\bm{p}}\cdot\hat{\bm{r}})^3 \hat{\bm{r}} \Big) \\
    \bm{u}^{(2)}_{F,3P3d}(\bm{k}) &= -\frac{\nu}{2\mu k^2} (\hat{\bm{p}}\cdot\hat{\bm{k}})^2 e^{-\omega^2/4} \left( 1 + \frac{\omega^2}{4} + \frac{\omega^4}{8} \right) (\mathbb{I}k^2 - \bm{k}\bm{k})\cdot\hat{\bm{p}},
\end{aligned}
\end{equation}
where we have defined the following auxiliary functions:
\begin{equation}\label{eq:H_K_3d}
\begin{aligned}
    H_{3P3d}(a) &= \erfc(a) + \frac{2 a}{\sqrt{\pi}} e^{-a^2} (1+6a^2-4a^4) \\
    I_{3P3d}(a) &= -9\erfc(a) - \frac{6 a}{\sqrt{\pi}} e^{-a^2} (3 + 2a^2 - 4a^4) \\
    J_{3P3d}(a) &= -3\erfc(a) - \frac{2a}{\sqrt{\pi}}e^{-a^2} (3+2a^2+20a^4 - 8a^6) \\
    K_{3P3d}(a) &= 15\erfc(a) + \frac{2 a}{\sqrt{\pi}}e^{-a^2} (15 + 10a^2 + 4a^4 - 8a^6).
\end{aligned}
\end{equation}
\section{Doubly periodic $3d$ multipoles} \label{sec: q2d}

\noindent We continue in the $2P3d$ geometry where the system is doubly periodic in the $x$- and $y$-directions and unbounded in the $z$-direction, as visualized in Figure \ref{fig: array-geometry}a. The position of the $3d$ force singularity is taken to be at $z = 0$, with an arbitrary $3d$ orientation $\hat{\bm{p}}$. The summation over periodic images (Eq.~\eqref{eq: sum-periodic}) now runs over all images in the $xy$-plane, with $\bm{r}_n = L(n_x \hat{\bm{x}} + n_y \hat{\bm{y}})$. As the real-space term is short-ranged, it can be evaluated in the same way as the $3P3d$ expressions \cite{Pozrikidis1996Periodic, deGraaf2017Ewald}. The long-ranged part $\mathbf{\Psi}_d(\bm{r})$ is more involved and requires a partially periodic extension of Poisson's summation formula~\citep{Lindbo2012Quasi2D, Grzybowski2000Quasi2D}:
\begin{equation} \label{eq: Poisson-quasi2D}
    \sum_{\bm{r}_n} \mathbf{\Psi}_{3d}(\bm{r} - \bm{r}_n) = \frac{1}{2\pi L^2} \sum_{\bm{q}} e^{i\bm{q}\cdot\bm{r}} \int_{-\infty}^\infty \text{d}k_z\, \mathcal{F}_{3d}\{ \mathbf{\Psi}_{3d}(\bm{r}) \} e^{ik_z z},
\end{equation}
where $\bm{q} = \frac{2\pi}{L}(\lambda_x \hat{\bm{x}} + \lambda_y \hat{\bm{y}})$ is the component of the wavevector in the periodic plane and $k_z$ its component in the non-periodic direction, so that $\bm{k} = \bm{q} + k_z \hat{\bm{z}}$. As only two of the three dimensions are periodic, the Fourier-space summation over $\bm{q}$ runs only over wavevectors oriented within the $xy$-plane, and the expression retains a dependence on the non-periodic coordinate $z$. The evaluation of the long-ranged part relies on the following integral identity \cite{Gradshteyn207Table}:
\begin{equation} \label{eq: Phi}
    \int_0^{\infty} \text{d}x \frac{e^{-\beta x^2} \cos(a x)}{\gamma^2 + x^2} = \frac{\pi}{4\gamma} e^{\beta \gamma^2} \bigg( e^{\gamma a} \erfc\Big(\gamma\sqrt{\beta} + \frac{a}{2\sqrt{\beta}}\Big)
    + e^{-\gamma a} \erfc\Big(\gamma\sqrt{\beta} - \frac{a}{2\sqrt{\beta}}\Big)\bigg).
\end{equation}
The resulting $k_z$-integral over the Fourier transform of $\mathbf{\Psi}_{3d}(\bm{r})$ can be expressed in terms of partial derivatives of Eq.~\eqref{eq: Phi}. By subsequently expanding the tensor $\mathbb{I}k^2 - \bm{k}\bm{k}$ as 
\begin{equation}
    \mathbb{I}k^2 - \bm{k}\bm{k} = \mathbb{I}(q^2 + k_z^2) - q^2 \hat{\bm{q}}\hat{\bm{q}} - k_z^2 \hat{\bm{z}}\hat{\bm{z}} - q k_z \left( \hat{\bm{z}}\hat{\bm{q}} + \hat{\bm{q}}\hat{\bm{z}} \right),
\end{equation}
terms can be grouped based on their $k_z$-dependence and evaluated separately. After lengthy calculations, we arrive at the following expression for the $2P3d$ Stokeslet velocity field:
\begin{equation} \label{eq: Ewald-q2D-Stokeslet}
\begin{aligned}
    \bm{u}_{2P3d}^{(0)}(\bm{r}) &= \sum_{\bm{r}_n} \bm{u}^{(0)}_{R,2P3d}(\bm{r}-\bm{r}_n) - \bm{u}_{F,0}^{(0)}(\ell) \\
    &+ \frac{1}{L^2}\sum_{\bm{q}\neq\bm{0}} \frac{1}{q} \left( \bm{u}^{(0)}_{F,2P3d}(\omega,\ell) \cos(\bm{q}\cdot\bm{r}) + \tilde{\bm{u}}^{(0)}_{F,2P3d}(\omega,\ell) \sin(\bm{q}\cdot\bm{r}) \right) \\
    \bm{u}^{(0)}_{R,2P3d}(\bm{r}) &= \frac{F}{8\pi\mu} \Big( C_{3P3d}(\xi r) \frac{\mathbb{I}}{r}+ D_{3P3d}(\xi r) \frac{\bm{r}\bm{r}}{r^3}\Big) \cdot\hat{\bm{p}} \\
    \bm{u}^{(0)}_{F,2P3d}(\omega, \ell) &= \frac{F}{8\mu} \Big( 2L\big(\omega, \ell\big) \mathbb{I} - M\big(\omega, \ell\big) \hat{\bm{q}}\hat{\bm{q}} - N\big(\omega,\ell\big)\hat{\bm{z}}\hat{\bm{z}} \Big) \cdot\hat{\bm{p}} \\
    \tilde{\bm{u}}^{(0)}_{F,2P3d}(\omega, \ell) &= \frac{F}{8\mu} P\big(\omega, \ell\big) \Big( \hat{\bm{q}}\hat{\bm{z}} + \hat{\bm{z}}\hat{\bm{q}} \Big) \cdot\hat{\bm{p}},
\end{aligned}
\end{equation}
where we have defined the dimensionless coordinates $\ell \equiv \xi z$ and $\omega \equiv q/\xi$ as well as the  auxiliary functions
\begin{equation}
\begin{aligned}
    L(\omega, \ell) &= \frac{2\omega}{\sqrt{\pi}} e^{-(\omega^2+4\ell^2)/4}\left(1 + \frac{\omega^2}{4} - \ell^2 \right) + \erfc \left(\frac{\omega}{2} - \ell \right)e^{-\omega \ell} + \erfc \left( \frac{\omega}{2} + \ell \right) e^{\omega \ell} \\
    M(\omega,\ell) &= \frac{2\omega}{\sqrt{\pi}} e^{-(\omega^2+4\ell^2)/4}\left( 1 + \frac{\omega^2}{2} \right) + \erfc \left( \frac{\omega}{2} - \ell \right) e^{-\omega \ell}(1+\omega \ell) + \erfc \left( \frac{\omega}{2} + \ell \right)e^{\omega \ell} \left(1 - \omega \ell \right) \\
    N(\omega,\ell) &= \frac{2\omega}{\sqrt{\pi}} e^{-(\omega^2+4\ell^2)/4}(1 -2\ell^2) + \erfc \left( \frac{\omega}{2} - \ell \right) e^{-\omega \ell} \left( 1-\omega \ell \right) + \erfc \left( \frac{\omega}{2} + \ell \right) e^{\omega \ell}(1 + \omega \ell) \\
    P(\omega, \ell) &= \frac{2\omega}{\sqrt{\pi}} \omega \ell e^{-(\omega^2+4\ell^2)/4} + \omega \ell e^{-\omega \ell} \erfc \left( \frac{\omega}{2}-\ell \right) + \omega \ell e^{\omega \ell} \erfc \left( \frac{\omega}{2} + \ell \right),
\end{aligned}
\end{equation}
while $C_{3P3d}(\xi r)$ and $D_{3P3d}(\xi r)$ are the same as for the triply periodic Stokeslet (Eq.~\eqref{eq:C_D_3d}). We note that our expressions formally differ from the corresponding solution derived by Lindbo and Tornberg~\citep{Lindbo2011QuasiStokes}, which uses a different splitting function. Our expressions furthermore exclude the self-interaction term discussed in~\citep{Lindbo2011QuasiStokes}, which should only be included when measuring the flow field at the position of the singularity. 

As in the triply periodic case, the Fourier space contribution diverges as $q \rightarrow 0$, which motivates the omission of the zero-wavevector mode from the summation, though the physical interpretation of this term is more subtle in the doubly periodic case \cite{Lindbo2011QuasiStokes, Grzybowski2000Quasi2D}: The zero-wavevector mode describes the mean flow resulting from the net force induced by the Stokeslet, whose periodic images are now placed only within the $xy$-plane rather than in the entire $3d$ domain. As the mean flow is therefore no longer constant throughout the domain, the flow field for $z \neq 0$ should be augmented. 

Following~\citep{Lindbo2011QuasiStokes} we thus subtract a $z$-dependent term to correct for this spurious flow, obtained as the difference between the total (un-split) Oseen tensor and the Fourier-space term, \textit{i.e.}, $\lim_{\bm{q}\rightarrow\bm{0}} \left( \mathbf{S}_{3d}(\bm{q}) - \mathbf{\Psi}_{3d}(\bm{q}) \right)$, which in our setting leads to
\begin{equation} \label{eq: k=0-corr}
    \bm{u}_{F,0}^{(0)}(\ell) = \frac{F}{2\mu\xi L^2} \left( \frac{\ell^2}{\sqrt{\pi}} e^{-\ell^2} - |\ell| + \ell \erf(\ell) \right) \bm{p}_\parallel, 
\end{equation}
where $\bm{p}_\parallel = \left( \mathbb{I}-\hat{\bm{z}}\hat{\bm{z}} \right)\cdot\hat{\bm{p}}$ denotes the in-plane component of $\hat{\bm{p}}$. The term proportional to $-|\ell|$ in Eq.~\eqref{eq: k=0-corr} encodes the cancellation of the far-field mean flow induced by the net force from the periodic array of Stokeslets, and is physically akin to adding a constant counter-force density to the layer containing the Stokeslet. 

\subsection{Force dipole}

\noindent Next, we derive an expression for the $2P3d$ force dipole. Taking the derivative of the real-space term $\bm{u}_{R,2P3d}^{(0)}(\bm{r})$, this is performed in the same way as in the triply periodic case. The spatial dependence of the Fourier-space terms is however no longer contained within the phase factor and requires more care: Unless $\hat{\bm{p}}$ lies fully within the (periodic) $xy$-plane, the expression should be differentiated with respect to $z$, which includes the auxiliary functions. Separating the in-plane ($\bm{p}_\parallel$) and out-of-plane ($p_z\hat{\bm{z}}$) components of the dipole orientation, we write $(\hat{\bm{p}} \cdot \nabla) = (\bm{p}_\parallel \cdot\nabla) + p_z (\hat{\bm{z}} \cdot\nabla)$, such that the corresponding terms can be treated separately: 
\begin{equation} \label{eq: Ewald-q2D-dipole}
\begin{split}
    \bm{u}_{2P3d}^{(1)}(\bm{r}) &= \sum_{\bm{r}_n} \bm{u}^{(1)}_{R,2P3d}(\bm{r}-\bm{r}_n) + \bm{u}_{F,0}^{(1)}(\ell) \\
    &+ \frac{1}{L^2} \sum_{\bm{q}\neq\bm{0}} \left( \bm{u}^{(1)}_{F,2P3d}(\omega,\ell) \sin(\bm{q}\cdot\bm{r}) + \tilde{\bm{u}}^{(1)}_{F,2P3d}(\omega,\ell) \cos(\bm{q}\cdot\bm{r}) \right)\\
    \bm{u}^{(1)}_{R,2P3d}(\bm{r}) &= \frac{\kappa}{8\pi\mu}\frac{1}{r^2} \Big( E_{3P3d}(\xi r) (\hat{\bm{p}}\cdot\hat{\bm{r}})^2 \hat{\bm{r}} + F_{3P3d}(\xi r) (\hat{\bm{p}}\cdot\hat{\bm{r}})\hat{\bm{p}} - D_{3P3d}(\xi r) \hat{\bm{r}} \Big) \\
    \bm{u}^{(1)}_{F,2P3d}(\omega,\ell) &= \frac{\kappa}{8\mu} \Big[ (\bm{p}_\parallel \cdot \hat{\bm{q}}) \Big( 2L\big(\omega, \ell\big) \mathbb{I} - M\big(\omega, \ell\big) \hat{\bm{q}}\hat{\bm{q}} - N\big(\omega,\ell\big)\hat{\bm{z}}\hat{\bm{z}} \Big) \\
    &\quad\quad\quad -p_z P_z^{(1)}\big(\omega, \ell\big) \Big( \hat{\bm{q}}\hat{\bm{z}} + \hat{\bm{z}}\hat{\bm{q}} \Big) \Big] \cdot\hat{\bm{p}} \\
    \tilde{\bm{u}}^{(1)}_{F,2P3d}(\omega,\ell) &= \frac{\kappa}{8\mu} \Big[ -(\bm{p}_\parallel \cdot \hat{\bm{q}}) P\big(\omega, \ell\big) \Big( \hat{\bm{q}}\hat{\bm{z}} + \hat{\bm{z}}\hat{\bm{q}} \Big) \\
    &\quad\quad\quad + p_z \Big( 2L_z^{(1)}\big(\omega, \ell\big) \mathbb{I} - M_z^{(1)}\big(\omega, \ell\big) \hat{\bm{q}}\hat{\bm{q}} + N_z^{(1)}\big(\omega,\ell\big)\hat{\bm{z}}\hat{\bm{z}} \Big) \Big] \cdot\hat{\bm{p}},
\end{split}
\end{equation}
where $D_{3P3d}(\xi r)$, $E_{3P3d}(\xi r)$ and $F_{3P3d}(\xi r)$ are defined as in $3P3d$ (Eq.~\eqref{eq: Ewald-3D-stresslet}). We furthermore define the following auxiliary functions for the out-of-plane component of the Fourier-space term:
\begin{equation}
\begin{aligned}
    L_z^{(1)}(\omega, \ell) &= \frac{8\ell}{\sqrt{\pi}} e^{-(\omega^2+4\ell^2)/4} \left(1 + \frac{\omega^2}{8} - \frac{\ell^2}{2} \right) + \erfc \left( \frac{\omega}{2} - \ell \right)e^{-\omega \ell} - \erfc \left( \frac{\omega}{2} + \ell \right) e^{\omega \ell} \\
    M_z^{(1)}(\omega,\ell) &= \omega \ell\frac{2\omega}{\sqrt{\pi}} e^{-(\omega^2+4\ell^2)/4} + \omega \ell \erfc \left( \frac{\omega}{2} - \ell \right) e^{-\omega \ell} + \omega \ell \erfc \left( \frac{\omega}{2} + \ell \right)e^{\omega \ell} \\
    N_z^{(1)}(\omega,\ell) &= \frac{8\ell}{\sqrt{\pi}} e^{-(\omega^2+4\ell^2)/4}(\ell^2 - 2) + \erfc\left( \frac{\omega}{2} - \ell \right) e^{-\omega \ell}(\omega \ell - 2) + \erfc \left( \frac{\omega}{2} + \ell \right) e^{\omega \ell}(\omega \ell + 2) \\
    P_z^{(1)}(\omega, \ell) &= \frac{2\omega}{\sqrt{\pi}} e^{-(\omega^2+4\ell^2)/4}(1-2\ell^2) + (1 - \omega \ell) e^{-\omega \ell} \erfc \left( \frac{\omega}{2}-\ell \right) + (1 + \omega \ell) e^{\omega \ell} \erfc \left(\frac{\omega}{2} + \ell \right).
\end{aligned}
\end{equation}
The zero-wavevector correction to the dipolar flow can be obtained through differentiation of the Stokeslet correction (Eq.~\eqref{eq: k=0-corr}). Crucially, however, the term containing $|\ell|$, which encodes the homogeneous, in-plane counterforce, should be excluded from the calculation, since the system is now automatically force free and, therefore no cancelling counterforce is required. This can be seen directly by averaging the Stokes equation over the periodic cell: The $xy$-averaged in-plane velocity $\bar{\bm{u}}_\parallel(z)$ satisfies 
\begin{equation}
    \mu\,\bar{\bm{u}}''_\parallel(z) = -\bar{\bm{f}}_\parallel(z),
\end{equation}
where the prime denotes differentiation with respect to the argument, and the averaged $z$-component of the velocity vanishes by incompressibility. The averaged in-plane force density corresponding to the dipole is furthermore given by 
\begin{equation}
    \bar{\bm{f}}_\parallel(z) = -\frac{\kappa \delta'(z)}{L^2} \, p_z\, \bm{p}_\parallel \,.
\end{equation}
Solving for the velocity, requiring $\bar{\bm{u}}_\parallel$ to remain bounded, and choosing the frame in which the far-field velocities above and below the plane are equal and opposite gives
\begin{align}
\bar{\bm{u}}_\parallel(z) = \frac{\kappa}{2\mu L^2}\, \mathrm{sgn}(z) \, p_z \bm{p}_\parallel,
\end{align}
\textit{i.e.} a jump discontinuity of size $\kappa p_z \bm{p}_\parallel/(\mu L^2)$ in the tangential velocity across the plane of the dipoles, which is non-zero whenever the dipole orientation simultaneously has non-zero in-plane and out-of-plane components. Retaining the $|\ell|$ term from Eq.~\eqref{eq: k=0-corr} is equivalent to adding a uniform layer of counter-dipoles that artificially cancels this physical jump, and this term should therefore be discarded. With this omission, we obtain the following correction term to the dipole flow:
\begin{equation}\label{eq:k=0_dipole}
    \bm{u}_{F,0}^{(1)}(\ell) = \frac{\kappa}{2\mu L^2} \left( \frac{2 \ell}{\sqrt{\pi}} e^{-\ell^2} (2 - \ell^2) + \erf(\ell) \right) p_z \, \bm{p}_\parallel.
\end{equation}
Unlike the corresponding correction for the Stokeslet (Eq.~\eqref{eq: k=0-corr}), $\bm{u}_{F,0}^{(1)}$ now goes to a constant, non-zero value as $z \rightarrow \pm \infty$ by virtue of the $\erf(\ell)$ term, leading to an asymptotic tangential flow whenever $\hat{\bm{p}}$ is diagonal to the periodic plane. 
This effect was briefly discussed in~\citep{Lindbo2011QuasiStokes} and is a direct consequence of the $xy$-periodicity: When moving away from the dipole in the $z$-direction, the effective singularity structure is no longer that of a single point dipole, but of an infinite, periodic array of dipoles of equal magnitude, each with a flow field that decays as $z^{-2}$, which sum up to give a constant value of the tangential flow as $z \rightarrow \pm \infty$. Geometrically, this effect is analogous to the spatially constant electric field generated by a homogeneously charged, infinite surface. 

\subsection{Force quadrupole}

\noindent The quadrupolar flow can be obtained using the same splitting into in-plane and out-of-plane components as for the dipole; note, however, that we now obtain cross-terms of in-plane and out-of-plane components, since $(\hat{\bm{p}}\cdot\nabla)^2 = p_z^2(\hat{\bm{z}}\cdot\nabla)^2 + (\bm{p}_\parallel\cdot\nabla)^2 + 2 p_z(\hat{\bm{z}}\cdot\nabla)(\bm{p}_\parallel\cdot\nabla)$. After lengthy algebra, the quadrupole field is obtained as
\begin{equation} \label{eq: Ewald-q2D-quadrupole}
\begin{aligned}
    \bm{u}_{2P3d}^{(2)}(\bm{r}) &= \sum_{\bm{r}_n} \bm{u}^{(2)}_{R,2P3d}(\bm{r}-\bm{r}_n) - \bm{u}_{F,0}^{(2)}(\ell) \\
    &+ \frac{1}{L^2} \sum_{\bm{q}\neq \bm{0}} q \left(\bm{u}^{(2)}_{F,2P3d}(\omega,\ell) \cos(\bm{q}\cdot\bm{r}) + \tilde{\bm{u}}^{(2)}_{F,2P3d}(\omega,\ell) \sin(\bm{q}\cdot\bm{r})\right) \\
    \bm{u}^{(2)}_{R,2P3d}(\bm{r}) &= \frac{\nu}{16\pi\mu}\frac{1}{r^3} \Big( H_{3P3d}(\xi r) \hat{\bm{p}} + I_{3P3d}(\xi r) (\hat{\bm{p}}\cdot\hat{\bm{r}})\hat{\bm{r}} 
    \\& \quad\quad\quad
    + J_{3P3d}(\xi r) (\hat{\bm{p}}\cdot\hat{\bm{r}})^2\hat{\bm{p}} + K_{3P3d}(\xi r) (\hat{\bm{p}}\cdot\hat{\bm{r}})^3 \hat{\bm{r}} \Big) \\
    \bm{u}^{(2)}_{F,2P3d}(\omega,\ell) &= \frac{\nu}{16\mu}  \Big[ -(\bm{p}_\parallel\cdot\hat{\bm{q}})^2 \Big( 2L\big(\omega, \ell\big) \mathbb{I} - M\big(\omega, \ell\big) \hat{\bm{q}}\hat{\bm{q}} - N\big(\omega,\ell\big)\hat{\bm{z}}\hat{\bm{z}} \Big) \\
    &\quad\quad\quad + p_z^2 \Big( 2L_z^{(2)}\big(\omega, \ell\big) \mathbb{I} + M_z^{(2)}\big(\omega, \ell\big) \hat{\bm{q}}\hat{\bm{q}} + N_z^{(2)}\big(\omega,\ell\big)\hat{\bm{z}}\hat{\bm{z}} \Big) \\
    &\quad\quad\quad + 2 p_z (\bm{p}_\parallel \cdot\hat{\bm{q}}) P_z^{(1)}(\omega,\ell)\left(\hat{\bm{q}}\hat{\bm{z}} + \hat{\bm{z}}\hat{\bm{q}} \right) \Big] \cdot\hat{\bm{p}} \\
    \tilde{\bm{u}}^{(2)}_{F,2P3d}(\omega,\ell) &= \frac{\nu}{16\mu} \Big[ -(\bm{p}_\parallel\cdot\hat{\bm{q}})^2 P\big(\omega, \ell\big) \Big( \hat{\bm{q}}\hat{\bm{z}} + \hat{\bm{z}}\hat{\bm{q}} \Big) 
    + p_z^2 P_z^{(2)}\big(\omega, \ell\big) \Big( \hat{\bm{q}}\hat{\bm{z}} + \hat{\bm{z}}\hat{\bm{q}} \Big) \\
    &\quad\quad\quad + 2 p_z (\bm{p}_\parallel \cdot\hat{\bm{q}}) \Big( 2L_z^{(1)}\big(\omega, \ell\big) \mathbb{I} - M_z^{(1)}\big(\omega, \ell\big) \hat{\bm{q}}\hat{\bm{q}} + N_z^{(1)}\big(\omega,\ell\big)\hat{\bm{z}}\hat{\bm{z}}
    \Big) \Big] \cdot\hat{\bm{p}},
\end{aligned}
\end{equation}
with $H_{3P3d}$, $I_{3P3d}$, $J_{3P3d}$, and $K_{3P3d}$ defined in Eq.~\eqref{eq:H_K_3d}, and
\begin{equation}
\begin{aligned}
    L_z^{(2)}(\omega, \ell) &= \frac{1}{\sqrt{\pi}\omega} e^{-(\omega^2+4\ell^2)/4}(12 + \omega^2 - 4\ell^2)(2\ell^2-1) + \erfc \left(\frac{\omega}{2} - \ell \right)e^{-\omega \ell} + \erfc \left( \frac{\omega}{2} + \ell \right) e^{\omega \ell} \\
    M_z^{(2)}(\omega,\ell) &= \frac{2\omega}{\sqrt{\pi}} e^{-(\omega^2+4\ell^2)/4}(1 - 2\ell^2) + (1 - \omega \ell) \erfc\left(\frac{\omega}{2} - \ell\right)e^{-\omega \ell} + (1 + \omega \ell) \erfc\left(\frac{\omega}{2} + \ell\right)e^{\omega \ell} \\
    N_z^{(2)}(\omega,\ell) &= \frac{8}{\sqrt{\pi} \omega} e^{-(\omega^2+4\ell^2)/4}(3 - 7\ell^2 + 2\ell^4) - \erfc\left(\frac{\omega}{2} - \ell\right)e^{-\omega \ell}(3 - \omega \ell) \\
    &\quad - \erfc\left(\frac{\omega}{2} + \ell\right)e^{\omega \ell}(3 + \omega \ell) \\
    P_z^{(2)}(\omega, \ell) &= \frac{8 \ell}{\sqrt{\pi}} e^{-(\omega^2+4\ell^2)/4}(\ell^2 - 2) + (\omega \ell - 2) e^{-\omega \ell} \erfc\left(\frac{\omega}{2}-\ell\right) + (\omega \ell + 2) e^{\omega \ell} \erfc\left(\frac{\omega}{2} + \ell\right).
\end{aligned}
\end{equation}
The corresponding $k \rightarrow 0$ correction is given by differentiation of Eq.~\eqref{eq: k=0-corr} excluding the $-|\ell|$ term, yielding 
\begin{equation}
    \bm{u}_{F,0}^{(2)}(\ell) = \frac{\nu \xi }{4\mu L^2} \left( \frac{2}{\sqrt{\pi}} e^{-\ell^2} (3 - 7\ell^2 + 2\ell^4) \right) p_z^2 \, \bm{p}_\parallel .
\end{equation}

\section{Doubly periodic $2d$ multipoles} \label{sec: 2d}

\noindent We now turn to the fully 2-dimensional, $2P2d$ geometry, where flows are described by the $2d$ Stokes equation. We begin by re-expressing the splitting function $g_d$ in a form suitable for the $2d$ Stokeslet. As detailed in~\citep{afKlinteberg2017Ewald}, the real-space splitting function can be connected to the real-space representation of a Fourier-space screening function, $\gamma_d$, through the relation $\gamma_{2d}(r) = \frac{1}{8\pi} \nabla^4 g_{2d}(r)$. We use the screening function introduced by Hasimoto \cite{Hasimoto1959Ewald3D}
\begin{equation}
\gamma_{2d}(k) = e^{-k^2/(4\xi^2)} \left( 1 + \frac{k^2}{4\xi^2} \right),
\end{equation}
leading to the following expression for the $2d$ splitting function:
\begin{equation}
    g_{2d}(\xi r) = - \Big( \frac{r^2}{2} \mathrm{Ei}\left(-(\xi r)^2\right) + \frac{1}{2\xi ^2}e^{-(\xi r)^2} \Big),
\end{equation}
where 
\begin{equation}
    \mathrm{Ei}(x) = \int_{-\infty}^x \text{d}t\, \frac{e^{t}}{t}
\end{equation}
is the exponential integral. Using this splitting function, we begin by deriving an Ewald-summed expression for the $2P2d$ Stokeslet. Such a solution was previously derived by Van de Vorst~\cite{vandeVorst1996Ewald2D} using an identity of the Gamma function and re-formulated in implicit form in terms of a generating function by Pozrikidis~\cite{Pozrikidis1996Periodic}. Here, we instead derive the point-force solution using the more straightforward splitting approach. The resulting expression is equivalent to the previous ones, except for a minor typo in the Gaussian factor of the reciprocal-space term in~\citep{Pozrikidis1996Periodic}, which is given as $e^{-1/(4\omega^2)}$ instead of $e^{-\omega^2/4}$.

We first evaluate the real-space part $\mathbf{\Theta}_{2d}(\bm{r})$ of the splitting function $g_{2d}(\xi r)$, yielding
\begin{equation}
    \mathbf{\Theta}_{2d}(\bm{r}) = (\mathbb{I}\nabla^2 - \nabla\nabla) g_{2d}(\xi r) =- C_{2P2d}(\xi r) \mathbb{I} + D_{2P2d}(\xi r) \frac{\bm{r}\bm{r}}{r^2},
\end{equation}
where we have defined the auxiliary functions
\begin{equation}
\begin{aligned}
    C_{2P2d}(a) &= 2 e^{-a^2} + \mathrm{Ei}(-a^2) \\
    D_{2P2d}(a) &= 2 e^{-a^2}.
\end{aligned}
\end{equation}
Next, we evaluate the Fourier-space term, which gives 
\begin{equation}
    \mathbf{\Psi}_{2d}(\bm{k}) = \frac{8\pi}{k^4} (\mathbb{I}k^2 - \bm{k}\bm{k}) \left( 1 + \frac{\omega^2}{4} \right) e^{-\omega^2/4}.
\end{equation}
Combining both terms, we can express the velocity field of a doubly periodic array of $2d$ Stokeslets as 
\begin{equation} \label{eq: ur-2D-Ewald-Stokeslet}
\begin{aligned}
    \bm{u}_{2P2d}^{(0)}(\bm{r}) &= \sum_{\bm{r}_n} \bm{u}^{(0)}_{R,2P2d}(\bm{r}-\bm{r}_n) + \frac{1}{L^2} \sum_{\bm{k}\neq\bm{0}} \bm{u}^{(0)}_{F,2P2d}(\bm{k}) \cos(\bm{k}\cdot\bm{r}) \\
    \bm{u}^{(0)}_{R,2P2d}(\bm{r}) &= \frac{F}{8\pi\mu} \Big( - C_{2P2d}(\xi r) \mathbb{I} + D_{2P2d}(\xi r) \frac{\bm{r}\bm{r}}{r^2} \Big) \cdot\hat{\bm{p}} \\
    \bm{u}^{(0)}_{F,2P2d}(\bm{k}) &= \frac{F}{\mu k^4} e^{-\omega^2/4} \left( 1 + \frac{\omega^2}{4} \right) (\mathbb{I}k^2 - \bm{k}\bm{k} ) \cdot\hat{\bm{p}}.
\end{aligned}
\end{equation}
We note that, apart from the prefactor $1/L^d$ and the dimensionality of the matrices, the Fourier-space part $\bm{u}_{F,2P2d}^{(0)}(\bm{k})$ is nearly identical to the triply periodic one (Eq.~\eqref{eq: Ewald-3D-Stokeslet}), with the only difference stemming from the different choices of splitting function.

The $2P2d$ dipole flow is obtained by taking the $2d$ gradient of Eq.~\eqref{eq: ur-2D-Ewald-Stokeslet} according to Eq.~\eqref{eq: stresslet}, leading to:
\begin{equation} \label{eq: ur-2D-Ewald-dipole}
\begin{aligned}
    \bm{u}_{2P2d}^{(1)}(\bm{r}) &= \sum_{\bm{r}_n} \bm{u}^{(1)}_{R,2P2d}(\bm{r}-\bm{r}_n) + \frac{1}{L^2} \sum_{\bm{k}\neq \bm{0}} \bm{u}^{(1)}_{F,2P2d}(\bm{k})\sin(\bm{k}\cdot\bm{r}) \\
    \bm{u}^{(1)}_{R,2P2d}(\bm{r}) &= \frac{\kappa}{4\pi\mu}\frac{e^{-(\xi r)^2}}{r} \Big( E_{2P2d}(\xi r)(\hat{\bm{p}}\cdot\hat{\bm{r}})^2\hat{\bm{r}}- \hat{\bm{r}} - F_{2P2d}(\xi r)(\hat{\bm{p}}\cdot\hat{\bm{r}})\hat{\bm{p}} \Big) \\
    \bm{u}^{(1)}_{F,2P2d}(\bm{k}) &= \frac{\kappa}{\mu k^3} (\hat{\bm{p}}\cdot\hat{\bm{k}})e^{-\omega^2/4} \left( 1 + \frac{\omega^2}{4} \right) (\mathbb{I}k^2 - \bm{k}\bm{k})\cdot\hat{\bm{p}},
\end{aligned}
\end{equation}
where we have defined the following auxiliary functions:
\begin{equation}
\begin{aligned}
    E_{2P2d}(a) &= 2 + 2 a^2 \\
    F_{2P2d}(a) &= 2 a^2.
\end{aligned}
\end{equation}
Similarly, the quadrupole flow is obtained \textit{via} Eq.~\eqref{eq: quadrupole-gen}, yielding
\begin{equation} \label{eq: ur-Ewald-2D-quadrupole}
\begin{aligned}
    \bm{u}_{2P2d}^{(2)}(\bm{r}) &= \sum_{\bm{r}_n} \bm{u}^{(2)}_{R,2P2d}(\bm{r}-\bm{r}_n)+ \frac{1}{L^2} \sum_{\bm{k}\neq \bm{0}} \bm{u}^{(2)}_{F,2P2d}(\bm{k}) \cos(\bm{k}\cdot\bm{r}) \\
    \bm{u}^{(2)}_{R,2P2d}(\bm{r}) &= \frac{\nu}{8\pi\mu}\frac{e^{-(\xi r)^2}}{r^2} \Big( H_{2P2d}(\xi r) \hat{\bm{p}} + I_{2P2d}(\xi r) (\hat{\bm{p}}\cdot\hat{\bm{r}})\hat{\bm{r}} 
    \\& \quad\quad\quad\quad\quad\quad\quad + J_{2P2d}(\xi r) (\hat{\bm{p}}\cdot\hat{\bm{r}})^2\hat{\bm{p}} + K_{2P2d}(\xi r) (\hat{\bm{p}}\cdot\hat{\bm{r}})^3 \hat{\bm{r}} \Big) \\
    \bm{u}^{(2)}_{F,2P2d}(\bm{k}) &= -\frac{\nu}{2\mu k^2} (\hat{\bm{p}}\cdot\hat{\bm{k}})^2e^{-\omega^2/4} \left( 1 + \frac{\omega^2}{4} \right) (\mathbb{I}k^2 - \bm{k}\bm{k})\cdot\hat{\bm{p}},
\end{aligned}
\end{equation}
with auxiliary functions
\begin{equation}
\begin{aligned}
    H_{2P2d}(a) &= 1 + 2 a^2 \\
    I_{2P2d}(a) &= -6(1+a^2) \\
    J_{2P2d}(a) &= -2(1+a^2 + 2a^4) \\
    K_{2P2d}(a) &= 4(2 + 2a^2 + a^4).
\end{aligned}
\end{equation}
We note that the $2P2d$ auxiliary functions have much simpler functional forms than the corresponding $3P3d$ expressions presented in Section~\ref{sec: 3d}. Furthermore, the short-ranged nature of the expression is governed by the Gaussian $e^{-(\xi r)^2}$ rather than the more computationally expensive error function $\erfc(\xi r)$, which ought to significantly improve the computational performance in the $2P2d$ geometry. 

\section{Numerical verification} \label{sec: numerical}

\noindent In this Section, we numerically verify the accuracy of the Ewald expressions derived in the previous sections by comparing with numerical results from a D3Q15 ($3P3d$ and $2P3d$ geometries) or D2Q9 ($2P2d$ geometry) BGK lattice Boltzmann (LB) solver together with a regularised point-force implementation, as discussed in several previous papers \cite{Nash2008Singular,BardfalvySoftMatter}. For the fully periodic cases, we set the square ($2P2d$) and cubic ($3P3d$) box lengths to $L = 500$ and $L = 100$ lattice points, respectively, with periodic boundaries in all directions. In the $2P3d$ geometry, the simulation box is instead enclosed by no-slip walls in the $z$-direction, which is also extended by an order of magnitude compared to the periodic directions to emulate as closely as possible an unbounded system, leading to the box dimensions $L_x = L_y = 50$ and $L_z = 500$. Note that, since the LB method solves the discretised (Navier-)Stokes equation in the low-Re regime by using alternating collision and streaming steps, periodic boundaries are imposed automatically by formulating the problem on a periodic domain. An explicit summation in the Ewald sense is therefore not needed, and the LB results thus constitute independent verification of the derived expressions.

In all cases, we place a single multipole in the centre of the simulation box. The hydrodynamic singularities are represented as sets of stationary, off-lattice regularized point forces as described in Fig.~\ref{fig: array-geometry}b, with $l$ taken to be equal to the LB lattice spacing, $\delta L$; for distances significantly larger than $\delta L$, we thus expect the LB results to coincide with the Ewald-summed, point-multipole results, while for short separations compared to $\delta L$ they will deviate both due to the extended nature of the singularities and the short-range regularisation inherent in the LB point-force implementation. In LB units, we use the swimmer force $F = 0.001571$, which is small enough to ensure that effects of fluid inertia are negligible, together with the dynamic viscosity $\mu = 1/6$. 

Exerting a solitary point force to the fluid leads to a non-zero net force, which necessitates a counter-acting set of forces to avoid an unphysical acceleration of the fluid volume and to allow for comparison with analytical results. In the $3P3d$ and $2P2d$ cases, we achieve this counter-forcing by placing a grid of oppositely oriented forces across all grid points in the simulation domain, whose sum exactly cancels the central force. In the $2P3d$ geometry, the counter-forces are instead only placed within the periodic plane of the Stokeslet, in agreement with our physical interpretation of the correction term in Eq.~\eqref{eq: k=0-corr}. For dipoles and quadrupoles, no additional forces are introduced beyond the central singularity, since these systems are force-free by construction. 

\begin{figure}[htbp]
    \centering
    \includegraphics[width=0.95\textwidth]{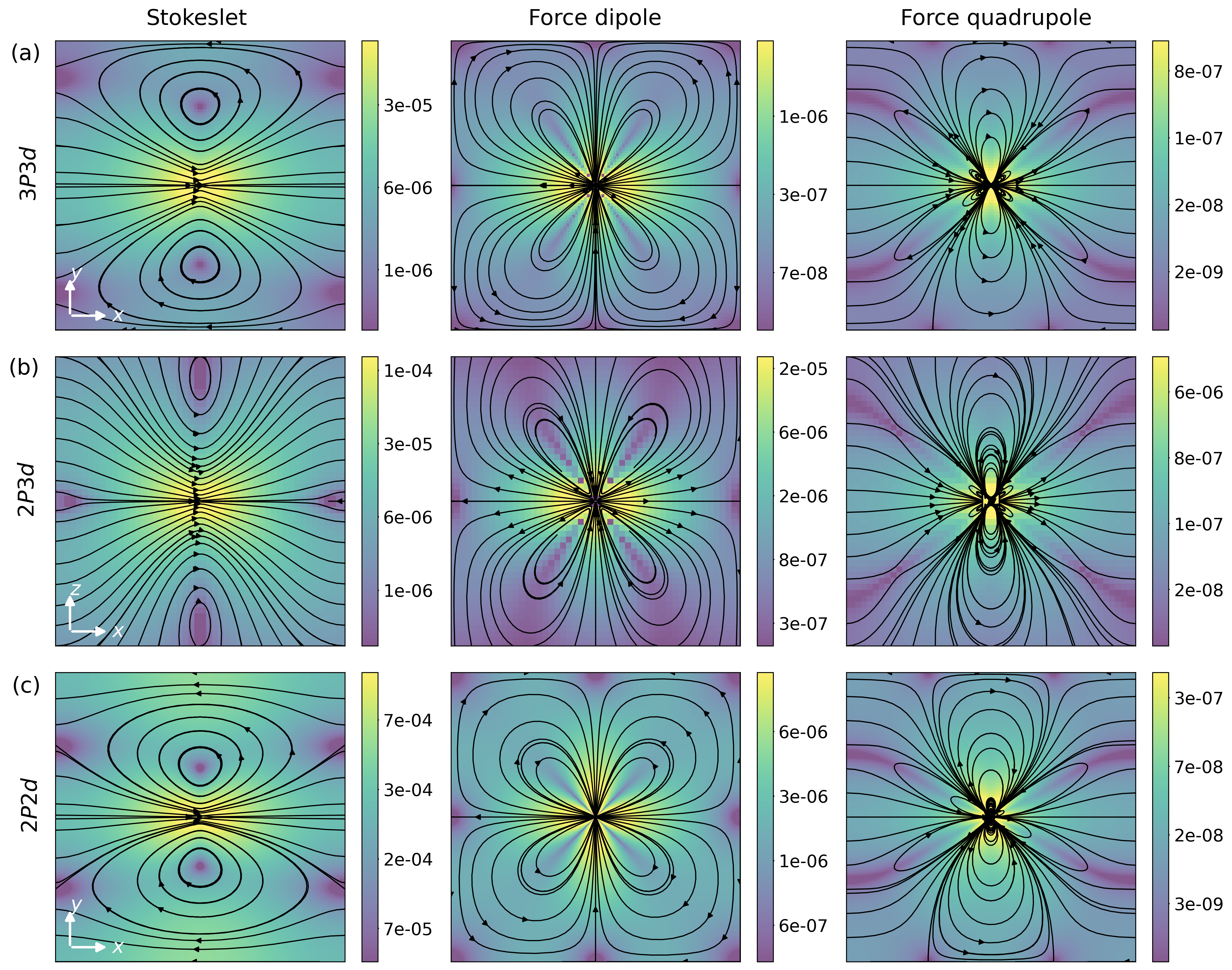}
    \caption{Streamlines and flow magnitudes of the Ewald-summed Stokeslet, force dipole and force quadrupole in (a) $3P3d$, (b) $2P3d$ and (c) $2P2d$ geometries. The singularities are oriented in the positive $x$-direction and the $3P3d$ and $2P3d$ results are shown in the $xy$- and $xz$-plane, respectively. The coloring indicates the flow magnitude on a logarithmic scale, with lighter colors corresponding to larger flow magnitudes.}
    \label{fig: sims-flow}
\end{figure}

\begin{figure}[htbp]
    \centering
    \includegraphics[width=0.95\textwidth]{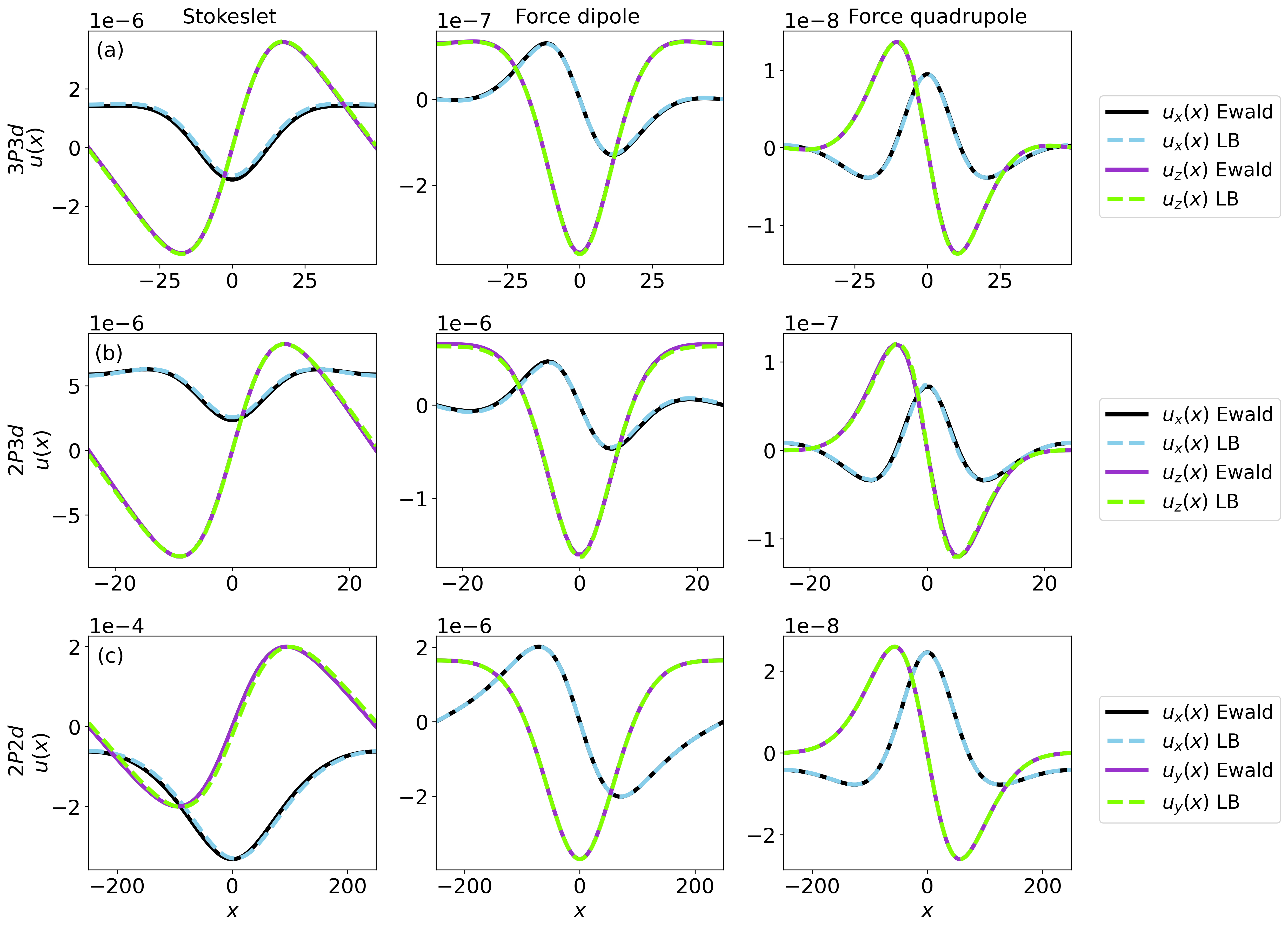}
    \caption{Cross-section of the flow velocity along the simulation box in the direction of the force for the Stokeslet (left), force dipole (centre) and force quadrupole (right) in (a) $3P3d$, (b) $2P3d$ and (c) $2P2d$ geometry. The cross-section is taken along the $x$-axis  with $y = 0$ and $z \approx L/4$ ($3P3d$), $y = 0$ and $z \approx L_x/4$ ($2P3d$) and $y \approx L/4$ ($2P2d$). The Ewald-summed results (solid lines) overlap almost completely with the LB results (dashed lines), apart from a small offset in the Stokeslet case resulting from the implementation of the background counterforce in the LB solver, verifying the validity of the former}
    \label{fig: sims-slices}
\end{figure}

\begin{figure}[htbp]
    \centering
    \includegraphics[width=0.95\textwidth]{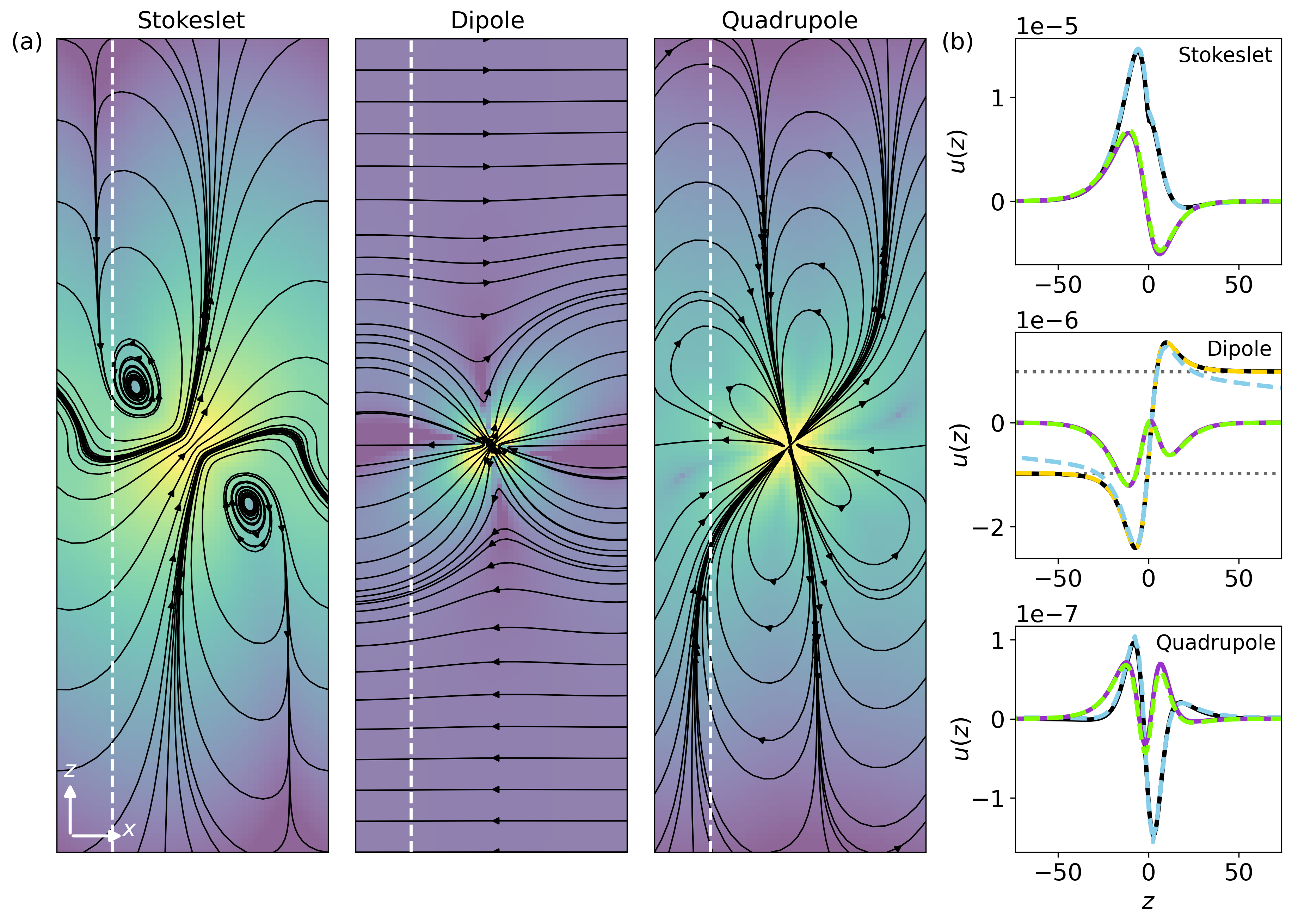}
    \caption{$2P3d$ flow fields for multipoles oriented diagonally in the $xz$-plane. (a) Streamlines of the Ewald-summed flow and (b) cross-sections of the velocity field for the Ewald expressions (solid lines) and LB results (dashed lines), measured along the white dotted lines in (a). Colors in (b) correspond to Fig.~\ref{fig: sims-slices}. In the centre plot of (b), the yellow dashed line denotes the flow velocity obtained through a direct evaluation of the lattice sum for the force dipole, and the grey dotted lines denote the asymptotic $z \rightarrow \pm \infty$ limits of Ewald-summed dipolar field}
    \label{fig: sims-flow-z}
\end{figure}

The corresponding Ewald expressions are evaluated numerically using the optimised value $\xi = \sqrt{\pi}/L$ of the splitting parameter~\citep{Beenakker1986Ewald} and the real-space and Fourier-space cutoffs $m_\text{max} = n_\text{max} = 10$, for which each Ewald expression has fully converged with a relative error within machine precision. The code used for generating the flow fields in Figs.~\ref{fig: sims-flow}--\ref{fig: sims-flow-z} can be accessed through a public GitHub repository~\cite{Code}. 

Fig.~\ref{fig: sims-flow} shows the Ewald-summed flow fields in each geometry: (a) $3P3d$, (b) $2P3d$, and (c) $2P2d$, showing the Stokeslet (left column), force dipole (centre column) and force quadrupole (right column) flows. The $2P3d$ flow fields (Fig.~\ref{fig: sims-flow}b) are shown in the non-periodic $xz$-plane, and differ from the $3P3d$ results through the lack of periodicity in the $z$-direction, leading to significantly more elongated recirculation loops than in the $3P3d$ case. The differences between the fully periodic $3P3d$ and $2P2d$ fields (Fig.~\ref{fig: sims-flow}c) instead result from the underlying features of 2-dimensional Stokes flows, which exhibit both slower radial decays and different angular dependencies compared to their 3-dimensional counterparts, where the latter differences are most apparent for the dipole and quadrupole fields.

In Fig.~\ref{fig: sims-slices} we quantitatively verify the Ewald expressions by measuring a cross-section of the velocity across the simulation box taken along the $x$-axis for each geometry and singularity, as described in the caption. For the dipole and quadrupole fields, the agreement is excellent across all geometries. For the Stokeslet, we observe a small discrepancy between the velocity profiles obtained through Ewald summation (solid lines) and with the LB method (dashed lines), which is due to a small shift of the LB flow field in the direction of the force, likely due to the approximate nature of the numerical force cancellation. 

As a final verification of our expressions, we turn to the non-trivial $2P3d$ case in which the multipole orientation points diagonally to the periodic plane, as visualised in Fig.~\ref{fig: sims-flow-z}a. Although the flow fields are strongly affected by this asymmetric configuration, the overall angular structures of the Stokeslet and quadrupole are preserved compared to the in-plane case. However, as discussed in conjunction with Eq.~\eqref{eq:k=0_dipole}, the dipole flow differs qualitatively from its fully-periodic counterpart in that its circulation loops now connect through the periodic ($x$) direction rather than looping back, which results in a constant (but opposite) tangential flow velocity as $z \rightarrow \pm \infty$, akin to a linear shear. As we confirm quantitatively in Fig.~\ref{fig: sims-flow-z}b, the Stokeslet and the quadrupole do not create any corresponding asymptotic net flow. For the Stokeslet, this is due to the cancellation of the net flow from the $k \rightarrow 0$ term, and for the quadrupole it follows from the fast $z^{-3}$ decay of the flow of the individual singularity, which decays to zero even when summed over two periodic directions. In all cases apart from the $x$-component of the dipole, we also get excellent agreement between LB results and the exact flow fields. In the dipolar case, the LB flow field (blue dashed lines) decays linearly when approaching the boundary of the domain due to the no-slip walls. To verify our Ewald results, we thus also compare to a direct, real-space evaluation of the lattice sum in its un-split form using summation in spherical shells (yellow dashed lines), which fully agrees with the Ewald expressions, yielding a final verification of the latter.

\section{Summary and outlook} \label{sec: conclusion}
\noindent In this Paper, we have derived and numerically verified analytical expressions for periodic arrays of hydrodynamic force multipoles in doubly periodic $2$- and $3$-dimensional geometries using Ewald summation. These results extend previous work on triply periodic Ewald-summed force dipoles~\citep{deGraaf2017Ewald} and provide expressions for the force dipole and force quadrupole in $2P3d$ (Eqs.~\eqref{eq: Ewald-q2D-dipole} and~\eqref{eq: Ewald-q2D-quadrupole}) and $2P2d$ (Eqs.~\eqref{eq: ur-2D-Ewald-dipole} and \eqref{eq: ur-Ewald-2D-quadrupole}) geometries, and furthermore provide new explicit expressions for the $2P3d$ Stokeslet using Beenakker's~\citep{Beenakker1986Ewald} splitting function (Eq.~\eqref{eq: Ewald-q2D-Stokeslet}). In particular, our derivation highlights some technical subtleties regarding the far-field correction for the dipole singularity in the $2P3d$ setting. 

Our work provides a natural extension of recent work on hydrodynamic interactions in $3d$ suspensions of spherical active particles with imposed surface slip flows (``squirmers'')~\cite{Deb2025EwaldColloid}. For a more accurate description of microswimmer flows in bulk $3d$ and quasi-$2d$ systems, the rotlet dipole singularity, which arises from the rotation and counterrotation of the bacterial body and flagella and decays as $1/r^3$, is also needed. This would naturally start from the previously derived expression of the Ewald-summed rotlet singularity ~\citep{afKlinteberg2016Rotlet}, although this singularity is primarily relevant in $3$-dimensional triply and doubly periodic geometries. For high-performance computational implementations, the expressions developed here can also be used as the basis for more computationally optimised Ewald representations of the Fourier-space term, as has previously been done for electrostatic lattice sums as well as for the periodic Stokeslet in $3P3d$ and $2P3d$ geometries~\citep{Darden1993PME, Toukmaji1996RevEwald, afKlinteberg2014Ewald, Lindbo2011QuasiStokes, Bagge2023StokesSpherical, Kosmacher2026StokesEfficient}.

\begin{acknowledgments}
\noindent 
For the purpose of open access, the author has applied a Creative Commons Attribution (CC BY) licence to any Author Accepted Manuscript version arising from this submission.
ZM and AM gratefully acknowledges financial support from the EPSRC (grant number EP/V048198/1). 
\end{acknowledgments}

\bibliography{refs_Ewald.bib} 

\end{document}